\documentclass[12pt,preprint]{aastex}
\newcommand{\nn}{\hphantom{0}}
\newcommand{\np}{\hphantom{.}}

\shorttitle{Pre-biotic Molecules in Arp~220}
\shortauthors{Salter et al.}
\begin{document}
\title{The Arecibo Arp~220 Spectral Census I: Discovery of the Pre-Biotic
Molecule Methanimine and New Cm-wavelength Transitions of Other Molecules}

\author{C.~J.~Salter, T.~Ghosh, B.~Catinella\altaffilmark{1},
        M.~Lebron\altaffilmark{2}, M.~S.~Lerner, R.~Minchin and 
        E.~Momjian\altaffilmark{3}}
\affil{Arecibo Observatory, HC~03 Box~53995, Arecibo, PR~00612}
\email{csalter@naic.edu}

\altaffiltext{1}{also, MPIfA, Karl-Schwarzschild-Stra{\ss}e 1, Postfach 1317,
D-85741, Garching, Germany}
\altaffiltext{2}{also, Department of Physical Sciences, University of
Puerto Rico, P.O. Box 23323, San Juan, PR 00931-3323 }
\altaffiltext{3}{also, NRAO, Array Operations Center, P.O. Box O, 1003
Lopezville Road, Socorro, NM 87801-0387}

\begin{abstract}

An on-going Arecibo line search between 1.1~and 10~GHz of the prototypical
starburst/megamaser galaxy, Arp~220, has revealed a spectrum rich in
molecular transitions. These include the ``pre-biotic'' molecules:
methanimine (CH$_{2}$NH) in emission, three $v_{2}=1$ direct l-type
absorption lines of HCN, and an absorption feature likely to be from
either $^{18}$OH or formic acid (HCOOH). In addition, we report the
detection of two, possibly three, transitions of $\lambda$4-cm excited OH
not previously detected in Arp~220 which are seen in absorption, and a
possible absorption feature from the 6.668-GHz line of methanol. This
marks the first distant extragalactic detection of methanimine,
a pre-biotic molecule.  Also, if confirmed, the possible methanol
absorption line presented here would represent the first extragalactic
detection of methanol at a distance further than 10~Mpc. In addition,
the strong, previously undetected, cm-wave HCN $v_{2}=1$ direct l-type
lines will aid the study of dense molecular gas and active star-forming
regions in this starburst galaxy.

\end{abstract}


\section{Introduction}

Star formation in galaxies occurs in two modes -- ``quiescent'', in
which stars form at a relatively modest rate over a long period of time
across the entire galaxy disk, and ``starburst'', in which gas is
turned into stars extremely rapidly, confined to compact regions 
of the galaxy, often its circumnuclear volume \citep{ken98}.
Starbursts are often triggered by external dynamical disturbances such
as galaxy mergers. The  dust heating associated with these intense
bursts of star formation within giant molecular clouds can produce
hugely increased IR luminosity and conditions favorable for maser
emission.  The strong 18-cm OH megamaser emission in some of these galaxies
is many orders of magnitude more luminous than its counterparts in our
Galaxy \citep{baan82,baan89,mar89,lons98,dar02}.  Ultra Luminous
Infrared Galaxies (ULIRGs) are thought to be systems where all of these
processes are occurring simultaneously \citep{els85,law86,sand96}.  The
existence of obscured AGNs in these objects are also considered as an
alternative (or additional) energy source where these are fueled by
molecular gas falling into the central regions of the merging systems
\citep{sur99,sco00,sand96,gen98,lutz99,vei02,vei06,arm07,iman07a}.  
Molecular gas is
thus one of the most important constituents of the ISM and plays a
critical role in the evolution of galaxies.

To date, over 140~molecules have been identified in space, mostly in the
ISM of our Milky Way Galaxy. Some of these are rather complex, containing
as many as 8~to 13~atoms.  Interstellar organic molecules are thought
to form mostly on the surface of dust grains. Heating events, such
as the formation of a protostar, release the icy grain mantles into
the gas phase \citep{nomu04,case93}.  Once released, these molecules
may form amino acids by the combination of organic species known as
``pre-biotic'' molecules \citep{blag03}.  \citep[We note that non-gas
phase reaction pathways for the formation of extraterrestrial amino acids
have been described in][]{elsi07}.  Methanimine is one such molecule
\citep[CH$_{2}$NH;][]{kirc73} which can form the simplest amino acid,
glycine (NH$_{2}$CH$_{2}$COOH), either by (i) first combining with
hydrogen cyanide (HCN) to form aminoacetonitrile (NH$_{2}$CH$_{2}$CN),
with subsequent hydrolysis \citep[Strecker Synthesis;] [] {dickr78},
or (ii) directly combining with formic acid (HCOOH) \citep{god73}.
Methanimine has been previously detected in the interstellar medium of our
own Galaxy \citep{god73,dickn97} and tentatively in the nearby galaxy,
NGC~253 \citep{mart06}, but never beyond the neighborhood of our Galaxy
(i.e. beyond $\sim$5~Mpc).

The majority of interstellar molecules (both galactic and extragalactic) 
have been discovered  at millimetric wavelengths, as molecules with
small moments of inertia are the most abundant cosmically, with their
rotational lines occurring at mm or shorter wavelengths. However,
although less abundant, many complex molecules have spectral lines
in the radio regime for $\lambda > 3$~cm, where ``line confusion''
does not set a limit to their detectability. Many transitions of small
polycyclic aromatic hydrocarbons (PAHs), pre-biotic molecules, and
even a number of transitions of the simplest amino acid, glycine, fall
within the relatively unexplored spectral range between 1~and 10~GHz.
In addition, observations in this frequency range are complementary to
mm spectral-line surveys which probe colder, lower density gas.

At Arecibo, we are conducting a spectral line census of Arp~220 between
1.1~and 10~GHz, for which the initial observations took place between 31
March and 22 April, 2007. It is planned that the remaining observations
will be made in 2008. Here we report the discovery of methanimine
emission in this galaxy, plus the detection of three $v_{2}=1$ direct
l-type absorption lines of HCN \citep{thor03} from the J=4, 5 and 6
vibrational levels.  Also reported is the possible detection of another
pre-biotic molecule, formic acid, albeit not unambiguously due to the
presence of a nearby $^{18}$OH line. In addition, we present the first
detections of two, possibly three, $\lambda$4-cm transitions of the
OH radical in absorption, high signal-to-noise ratio detections of the
$\lambda$6- and 5-cm transitions of OH, and the possible detection of
the 6.7-GHz methanol (CH$_{3}$OH) transition in absorption.

\section{Arp~220}

At a distance of $\sim$77\,Mpc (redshift, z=0.018126), Arp~220 is the
nearest Ultra-Luminous Infra-Red Galaxy. Much of its IR luminosity arises
from a powerful, dust-enshrouded starburst, triggered by the merger of
two gas-rich galaxies \citep{sak99, mun01, sand96}.  Evidence for this is
provided by high resolution optical and radio images revealing a double
nucleus with tidal tails and dust lanes. A high supernova rate has been
found from recent high resolution VLBI studies \citep{lons06}.  Molecules
such as the OH radical \citep{baan82, ghosh03}, CO \citep{sco86},
formaldehyde \citep{ara04}, ammonia \citep{taka05}, and mm-transitions
of HCN \citep{sol92,evan06,iman07b} have been detected in Arp~220.
In fact, \citet{dow07} have recently used high resolution imaging of 
the CO(2--1) line and the $\lambda$1.3~mm dust radiation to
provide strong evidence for the existence of a ``buried'' AGN in the
western nucleus of the galaxy, a conclusion heavily affecting
interpretation of the situation within Arp~220.

Arp~220 is also known as the prototype OH megamaser galaxy. Its OH
$\lambda$18-cm maser emission was first reported by \citet{baan82}.
High-resolution maps of the OH maser revealed complex structures
which could be interpreted either as a dual-component distribution
\citep{diam89,lons98,rovi03} or, by clumpy unsaturated masers within a
single-component medium \citep{PCEP05,MOM06}.

\section{Observations}

Using several of the complement of receivers on the Arecibo 305-m
telescope, we are in the process of making an almost-complete spectral
scan of Arp~220 between 1.1~and 10~GHz. To do this, we have employed
the WAPP (Wideband Arecibo Pulsar Processor) spectrometer in its
recently-commissioned ``dual-board" mode. In this mode, eight independent
boards, each of 100-MHz bandwidth with 3-level quantization, can be used
to cover a spectral band of up to 800~MHz at a single time.  Compared to
the earlier WAPP capacity, this doubles the total bandwidth covered.
The present project serves both for the scientific commissioning of this
new option, and as a demonstration of its capabilities.

In practice, we have overlapped the eight boards for each observation
such that their centers are separated by 85~MHz. As the final 5-or-so
MHz are affected by filter roll-off, this allows high quality data to
be acquired for an instantaneous bandwidth of 680~MHz, meaning that
those Arecibo receivers which have total bandwidths of 2~GHz can be
fully covered via three separate frequency settings. The basic spectral
resolution is 24.4~kHz, and both orthogonal polarizations of the
celestial signal are recorded.

The observations are being made via a modified version of the
Double Position Switching (DPS) technique \citep{ghos02}. An ON/OFF
position-switched observation with 5-min component phases is made
on Arp~220, followed by an ON/OFF with 1-min phases on the strong,
angularly-nearby, continuum source, J1531+2402, which is used as
a band-pass calibrator. In the presence of band-pass ripples or
trapped-modes in the orthomode transducer of the feed, this strategy is
necessary at all frequencies to produce acceptable spectral baselines.
In addition, apparent features observed in the data can often be
recognized as astronomical, and not due to the presence of radio frequency
interference (RFI), via a comparison of the spectra for Arp~220 and
J1531+2402.  Data for a total ON-source observing time of about 60~min
for Arp~220 will eventually be acquired for each frequency setting.

Data reduction has been performed using the Arecibo IDL analysis
package written by Phil Perillat. The individual ON/OFF scans on Arp~220
were processed to yield (ON-OFF)/OFF spectra, and these were ``bandpass
corrected'' using similar spectra for J1531+2402. Each individual Arp~220
and J1531+2402 scan was inspected for quality, and all RFI present was
noted. A number of scans were rejected either due to technical problems
or because of excessive RFI. All acceptable scans for a particular
frequency setting were then co-added to produce the final spectra.
These spectra were smoothed in frequency to a number of resolutions,
and the resultant spectra inspected visually to identify the presence
of possible emission or absorption lines.  Considerable cross-checking
was performed to ensure that candidate lines were real, and not the
effect of RFI or equipment problems.

\section{Results \& Discussion}

A complete cm-wavelength line census for Arp~220, and its full
implications for the physical and chemical properties of the
interstellar medium of this galaxy, are deferred until completion of
the observations. Here we present the detection of methanimine in
emission, and of the $v_{2}=1$ direct l-type transitions of HCN,
excited-OH transitions, a line that could be either $^{18}$OH or
HCOOH, and possibly methanol in absorption for this prototypical
nearby ULIRG. The spectra are presented in Figs.~\ref{f1}--\ref{f8},
and the parameters derived from them are given in Tables~1 \&~2. 

Table~1 presents the results for the emission line of methanimine.
The table is ordered as follows: Columns (1--3) present the
name of the molecule, the relevant transition and its rest frequency
in MHz. Col.~(4) is the peak flux density of the line in mJy. Col.~(5)
is the rms noise on the spectrum in mJy at a velocity resolution of
$\sim$30~km\,s$^{-1}$.  Col.~(6) is the radial velocity in km\,s$^{-1}$
at the peak of the measured line, derived from the rest frequency
of the strongest of the six lines in the multiplet (i.e. that at
5289.813~MHz). Col.~(7) is the full-velocity width in km\,s$^{-1}$
of the entire emission feature at half of the peak intensity.

Table~2 presents the results for lines seen in absorption. The
table is ordered as follows: Columns (1--3) are as for Table~1. Col.~(4)
is the maximum optical depth of the absorption line. Col.~(5) is the
rms noise on the optical-depth spectrum at a velocity resolution of
$\sim$30~km\,s$^{-1}$. Col.~(6) is the radial velocity in km\,s$^{-1}$
at the maximum optical depth. Col.~(7) is the full width in km\,s$^{-1}$
of the line at half of the peak optical depth. Col.~(8) is $\int \tau dV$,
the integrated area in km\,s$^{-1}$ under the absorption feature.

In the following sections, we discuss the spectra of each molecule
separately.

\subsection{\it Methanimine (CH$_{2}$NH)}

\begin{figure} 
\epsscale{0.7}
\plotone{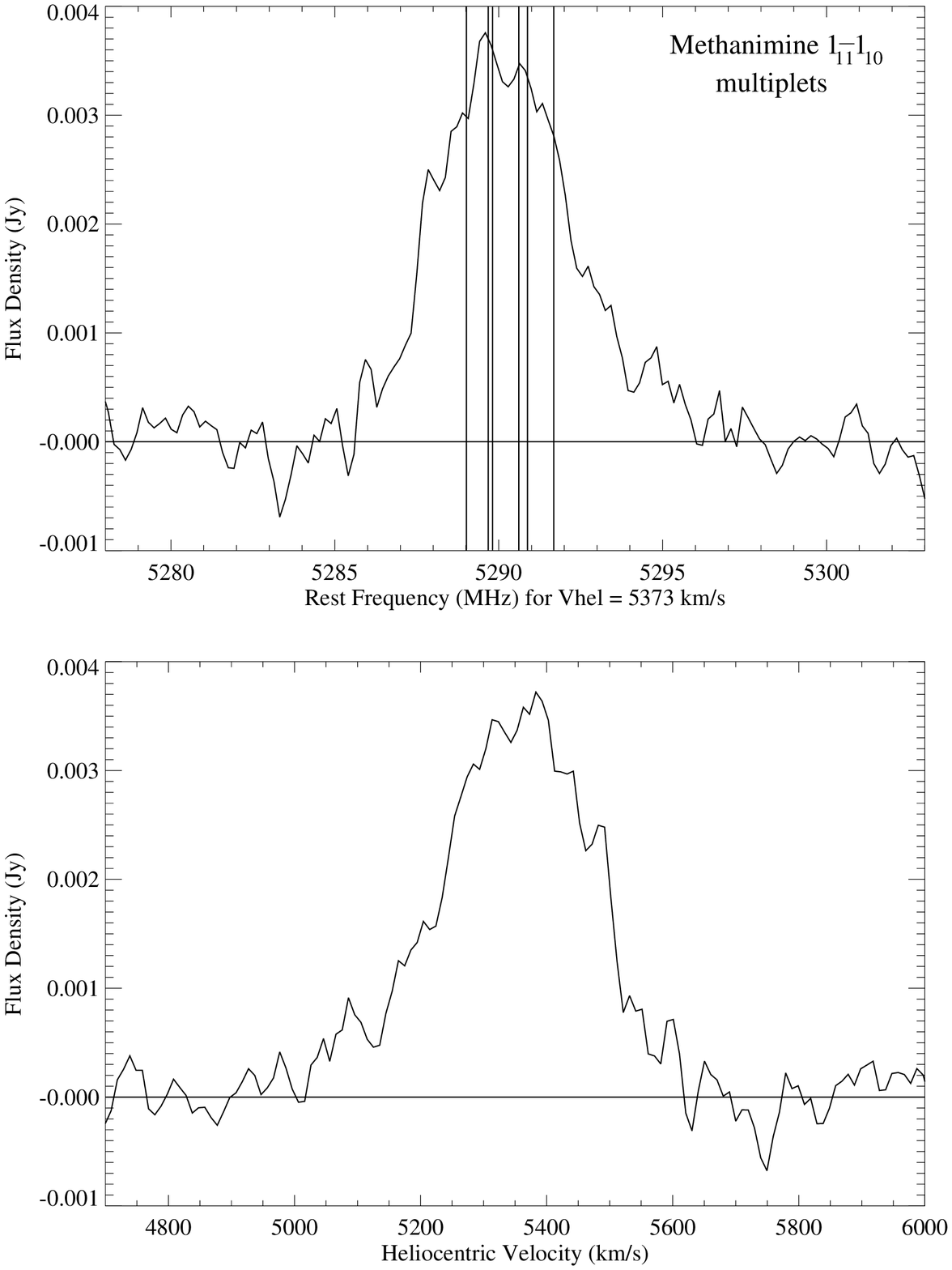}
\caption{The blended emission line from the six $1_{10}$ -- $1_{11}$
multiplet transitions of methanimine in Arp~220.  The rest frequencies of
the individual transitions are shown by vertical lines in the upper panel
for an assumed heliocentric velocity of 5373~km\,s$^{-1}$ (corresponding
to the western nucleus of the galaxy). The lower panel shows the same
spectrum as a function of the heliocentric velocity appropriate for the
strongest line (rest frequency = 5289.813~MHz) in the multiplet. The
velocity resolution is $\sim$30~km\,s$^{-1}$.}
\label{f1}
\end{figure}

\begin{figure}
\centerline{\resizebox{\columnwidth}{!}{\rotatebox{270}{
\includegraphics{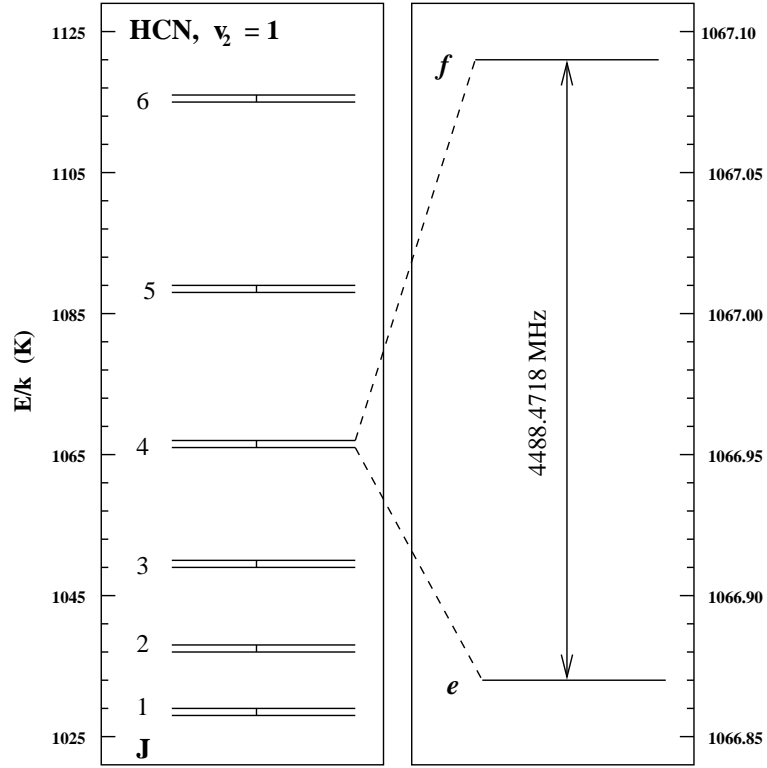}}}}
\caption{Term value diagram for HCN in its $v_{2}$=1 vibrational state
from $J=1$ to $J=6$.  Only the direct {\it l-type} transitions with
$\Delta J=0$ are shown.  The diagram on the right-hand side shows the
$J=4$ direct {\it l-type} transition at 4488.4718~MHz in detail. We
have presently observed the $J = 2, 4, 5$ and 6 transitions at 1346,
4488, 6731 and 9423~MHz respectively (see Fig.~\ref{f2}). This diagram
was drawn using the energy level values obtained from the CDMS database
\citep{mul05,mul01}}.
\label{hcn-engdia}
\end{figure}

\begin{figure}
\epsscale{1.0}
\plotone{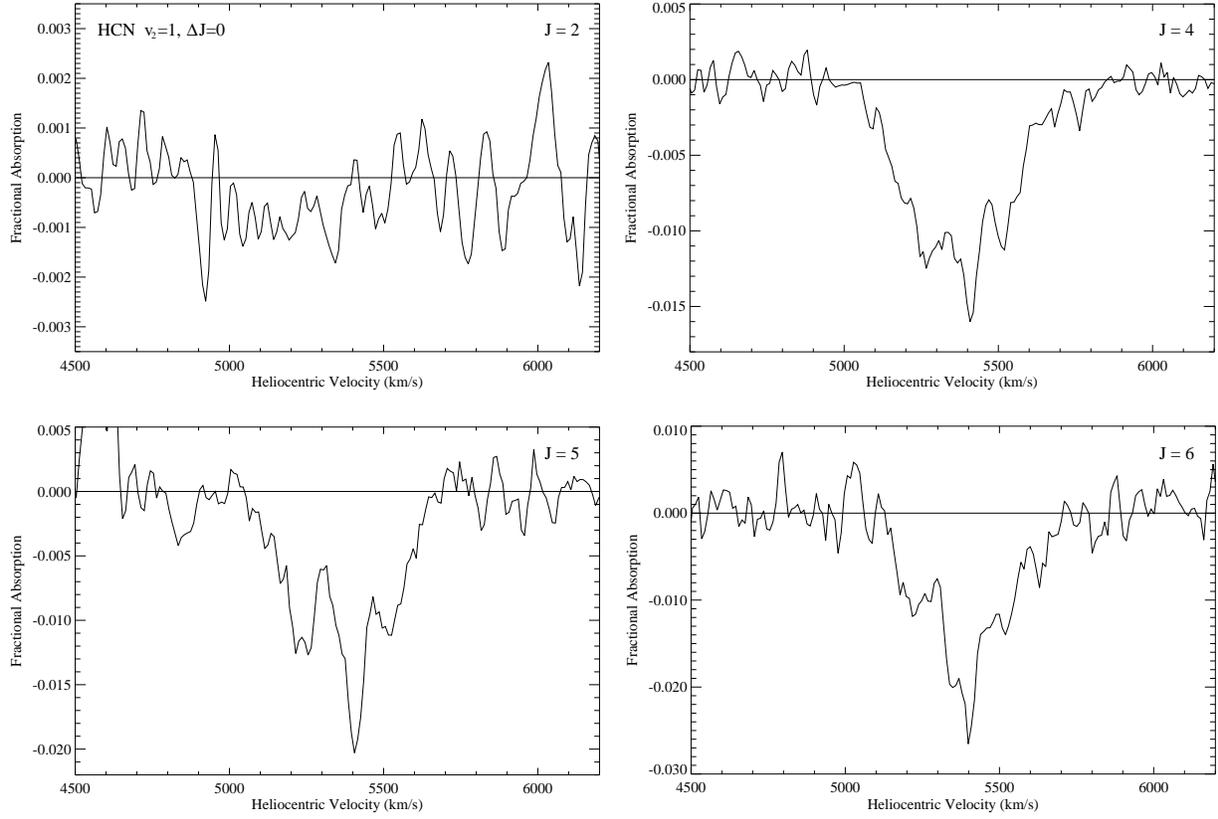}
\caption{The first astronomical detections of the $v_{2}=1$ direct
l-type absorption lines of HCN with vibrational levels J=4, 5 and~6
(at 4488, 6731 and 9423~MHz respectively). The spectra are plotted
with heliocentric velocity as abscissa. The non-detection of the J=2
vibrational level (at 1346~MHz) is also included in the figure. The
velocity resolution is $\sim$30~km\,s$^{-1}$.
\label{f2}}                                                                 
\end{figure}

The  broad emission feature covering all six $1_{10}$ -- $1_{11}$ transitions
of the C-band multiplet of methanimine is presented in Fig.~\ref{f1}.
In Fig.~\ref{f1}a, the rest frequencies of the transitions in the frame of
Arp~220 were derived using a recessional velocity of 5373~km\,s$^{-1}$,
as found for the western component of the  OH megamaser emission of
Arp~220, while the heliocentric velocity axis of Fig.~\ref{f1}b is
for the transition expected to be the strongest in the methanimine
multiplet, that at a rest frequency of 5289.813~MHz \citep{god73}. The
total velocity width (FWHM) of this emission line is 270~km\,s$^{-1}$;
we note that a large velocity width is observed for almost all molecular
spectra in Arp~220 \citep[e.g.] [] {taka05}.  Using the angular
size of $0.27^{\prime\prime} \times 0.21^{\prime\prime}$ measured
for the strongest component of the formaldehyde (H$_{2}$CO) emission
\citep{baan95} as an upper limit for the angular size of the methanimine
emission region, we derive a lower limit to the brightness temperature
of $\sim$2800~K.  Taking into account the six lines in the multiplet and
their relative intensities,  we calculate a reduced value of this lower
limit to be $\sim$1000~K for the strongest component. This  is similar
to the methanimine decomposition temperature of 1300~K \citep{nguy96}.
If the solid angle of the emission is smaller than the above, for instance
if the emission comes from a number of very compact components, then
the brightness temperature could be much higher.  We conclude that,
as for formaldehyde \citep{ara04}, methanimine in Arp~220 is likely to
be showing weak maser emission for this transition.  We have been
awarded time with the MERLIN array to observe this source at higher
spatial resolution in order to determine the brightness temperature
more accurately and to discover where exactly in the galaxy the emission
is located.

\subsection{\it Hydrogen Cyanide (HCN)}

Fig.~\ref{hcn-engdia} shows the energy diagram for the $v_{2}$=1
direct l-type transitions of hydrogen cyanide (HCN) in the J=1 to J=6
vibrational levels.  Our spectra of the J=2, 4, 5 and 6 HCN transitions
are presented in Fig.~\ref{f2}.  The J=4, 5 and 6 transitions are detected
in absorption against the continuum emission of Arp~220, and we place an
upper limit at the 3-$\sigma$ level of 0.0025 for the optical depth of
the J=2 line.  The first detailed study of this type of HCN transition
was carried out for the Galactic proto-planetary nebula, CRL\,618,
by \citet{thor03}. These observers detected a number of the higher
vibrational levels in this HCN ladder (J=8--14). However we note that
the lower energy transitions presented here seem not to have been
previously detected in any celestial source.  We also note that these
lines represent a high excitation energy above the HCN ground state,
(e.g. 1067~K for the J=4 line.)

HCN is a well known indicator of high gas density.  
\citet{gao04} demonstrated that a very strong linear correlation exists
between the luminosities L$_{\rm IR}$ and L$_{\rm HCN}$ for mm-wave
transitions of HCN, extending over a wide range of L$_{\rm IR}$
from normal galaxies to ULIRGs. They found the similar L$_{\rm IR}$ --
L$_{\rm CO}$ correlation to have a less linear form, marking out L$_{\rm
HCN}$ as the best tracer of dense molecular gas mass in galaxies, and
hence of active star-formation.  However, \citet{gc08} present evidence
that L$_{\rm IR}$/L$_{\rm HCN}$ is systematically higher in (U)LIRGs
than in normal star forming galaxies.  The relative line integrals from
the J=4 and 6 lines suggest an approximate excitation temperature of
$\sim 150$~K.  These transitions represent high excitation, and given
the complicated scenarios now emerging for the central region of Arp~220,
it would be of great interest to ascertain the precise circumstances in
which these absorptions arise.
% it would seem 
% plausible that they are coming from regions of star formation.

Unlike the relatively ``smooth'' line profiles seen for the excited-OH
lines (e.g.\ Fig.~\ref{f3}), the HCN lines each show evidence for the
presence of a number of discrete components. In fact, the central,
strongest HCN component becomes increasingly dominant from J=4 to 6.
The absence of the J=2 line requires explanation, as the predicted
absorption line is expected to be an order of magnitude stronger than
the 3-$\sigma$ limit we place above. It is highly improbable that the
fraction of the continuum emission ``covered'' by the clouds producing
the HCN absorption could have decreased by such a large amount between 4488 and
1347~MHz that the J=2 line is rendered unobservable. Much more likely is that
free-free absorption in the foreground ionized screen of this starburst
galaxy is greatly attenuating the J=2 line. For this to be the case,
an opacity of $\gtrsim 2.25$ at 1347~MHz (in the rest frame of Arp~220)
would be required. This would imply an optical depth of $\gtrsim 1.5$
at 1630~MHz in the galaxy, as found for the spectra of a number of the
SNRs near the twin nuclei of Arp~220 by \cite{par07}. These authors
attributed this high opacity to the combined presence of ``a patchy FFA
[free-free absorption] ISM with a median opacity of less than 1'',
plus absorption in the regions of ionized circumstellar mass-loss
envelopes surrounding the SN progenitors. The HCN lines we see are
expected to arise from star-forming regions of high gas density, and
a detailed study of all possible lines in the $v_{2}=1$ direct l-type
transitions of the HCN ladder would provide useful evidence concerning
the properties of the foreground gas screen to these regions. Clearly,
the spectrum of the HCN J=3 line (with a rest frequency of 2693.3~MHz)
will be crucial to such a study, and we will be observing this line
during our 2008 campaign to complete the present observations. It should
be noted that for the implied optical depths at 1347~MHz, even the J=4
line will be reduced through FFA by $\gtrsim$~20\%, which would reduce
the above derived excitation temperature to $\sim$120~K.

\subsection{\it Excited Hydroxyl (OH)}
\label{OH_lines}

\begin{figure}
\epsscale{0.8}
\plotone{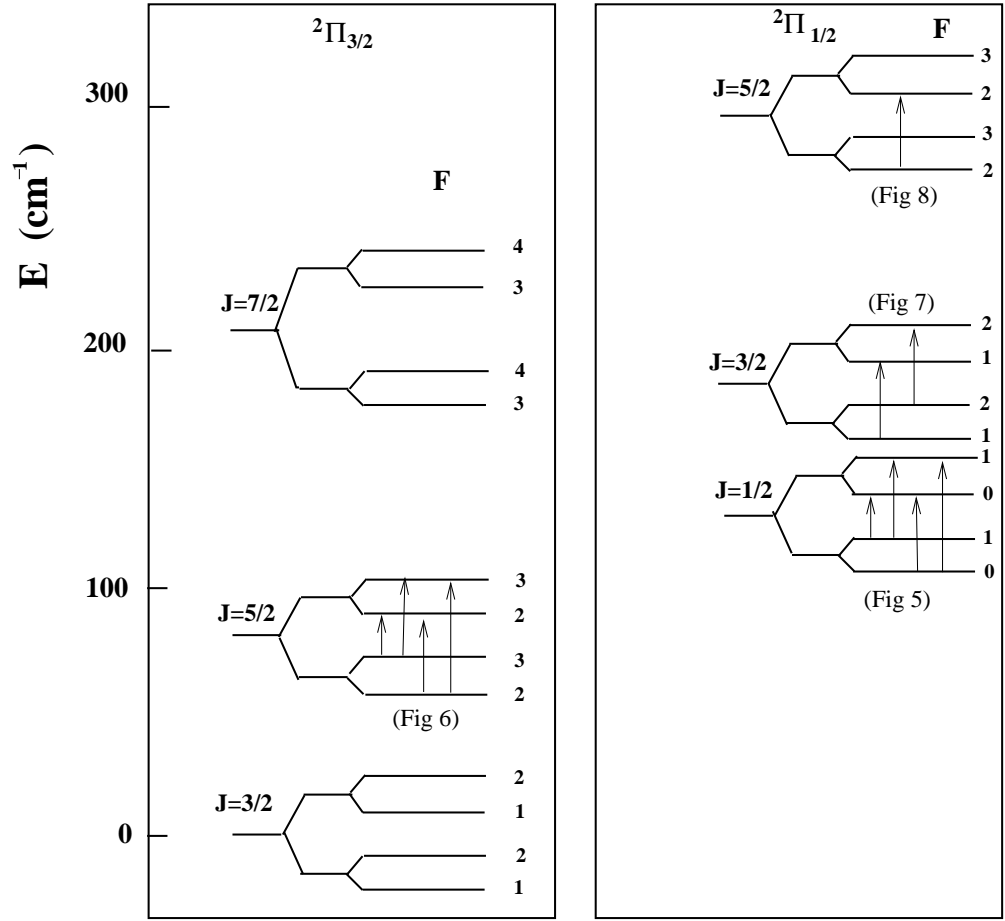}
\caption{The energy diagram (not to scale) for various $\Lambda$-doublet
transitions of the OH radical. The transitions that we have detected
(all in absorption) are shown by upward directed arrows, with the
relevant figure numbers also marked. We note that $2\Pi_{1/2}$, J=1/2,
F=0-0 and 1-1 have identical transition energies.}
\label{OH_energy_diag}
\end{figure}

\begin{figure}
\epsscale{0.6}
\plotone{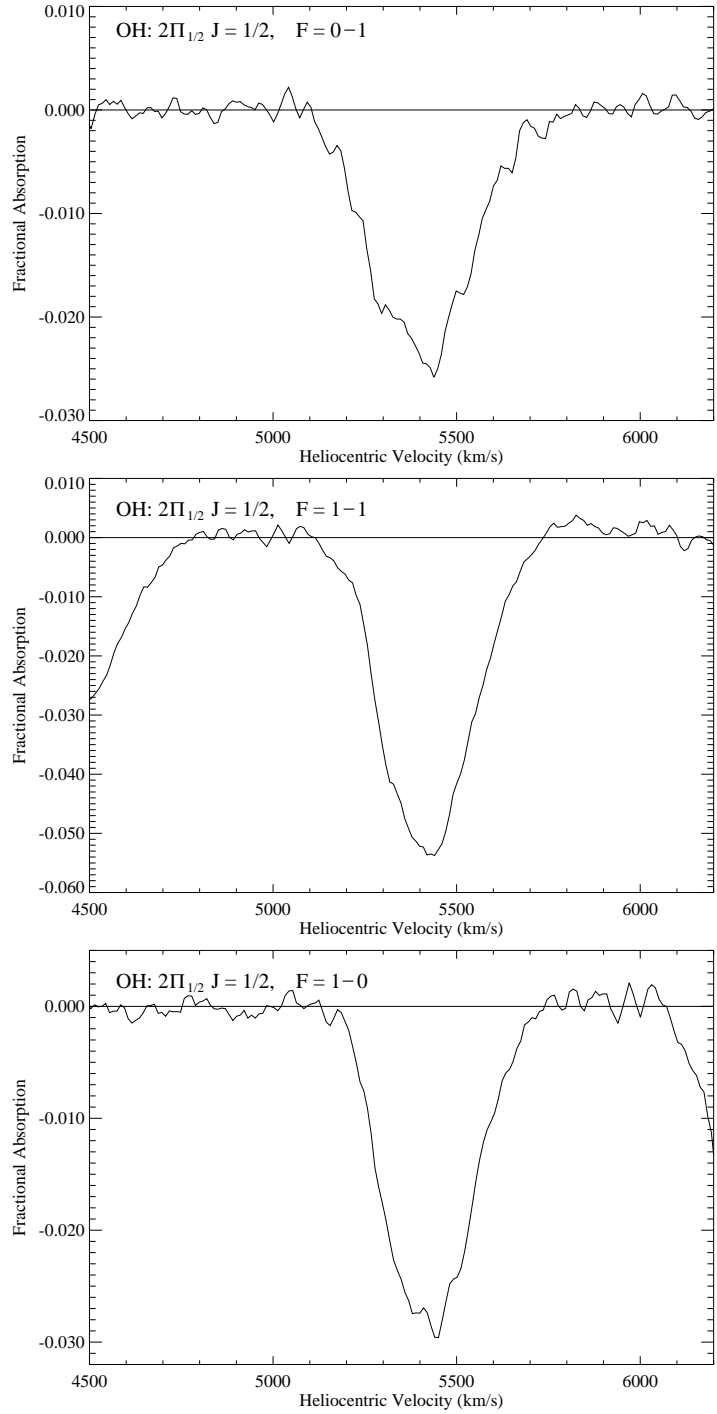}
\caption{The three $\lambda$6-cm excited OH transitions from
$2\Pi_{1/2}$, J=1/2, F=0-1, 1-1/0-0 and 1-0 (at 4660, 4750 and 4765~MHz
respectively). The spectra are plotted with heliocentric velocity as
abscissa.  The F=1-1/0-0 line is about twice as strong as the other two
lines. The velocity resolution is $\sim$30~km\,s$^{-1}$.
\label{f3}}                                                                 
\end{figure}

\begin{figure}
\epsscale{0.8}
\plotone{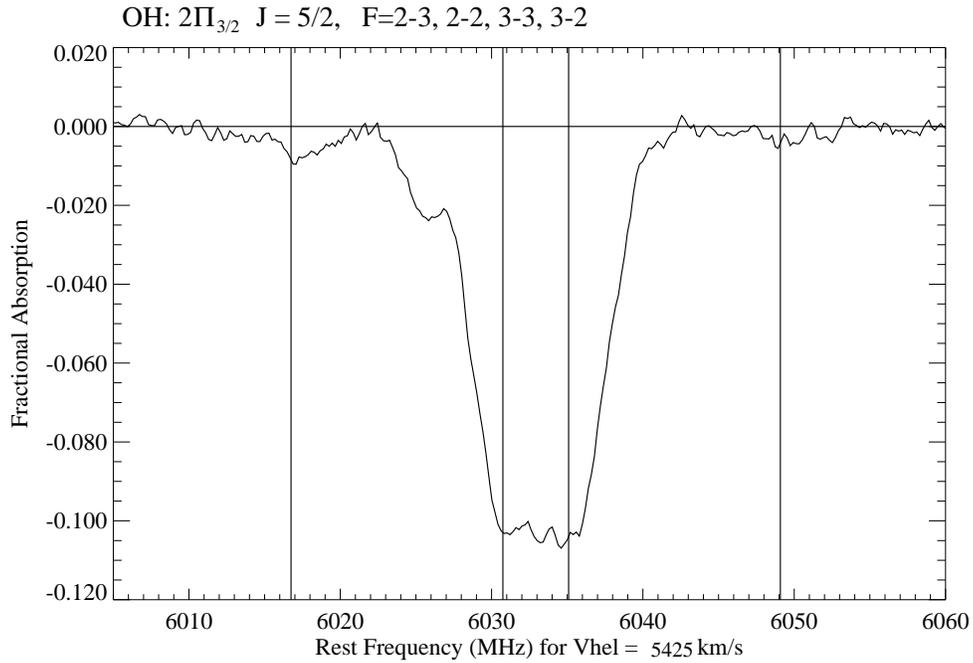}
\caption{The four $\lambda$5-cm excited OH transitions as a function of
rest frequency for an assumed heliocentric velocity of 5425~km\,$s^{-1}$
(as appropriate for the $\lambda$6-cm OH lines). The frequencies of
the four transitions $2\Pi_{3/2}$, J=5/2, F=2-3, 2-2, 3-3 and 3-2 are
indicated by vertical lines. The two main lines form a blended absorption
line profile. The velocity resolution is $\sim$30~km\,s$^{-1}$.
\label{f4}}                                                                 
\end{figure}

\begin{figure}
\epsscale{0.7}
\plotone{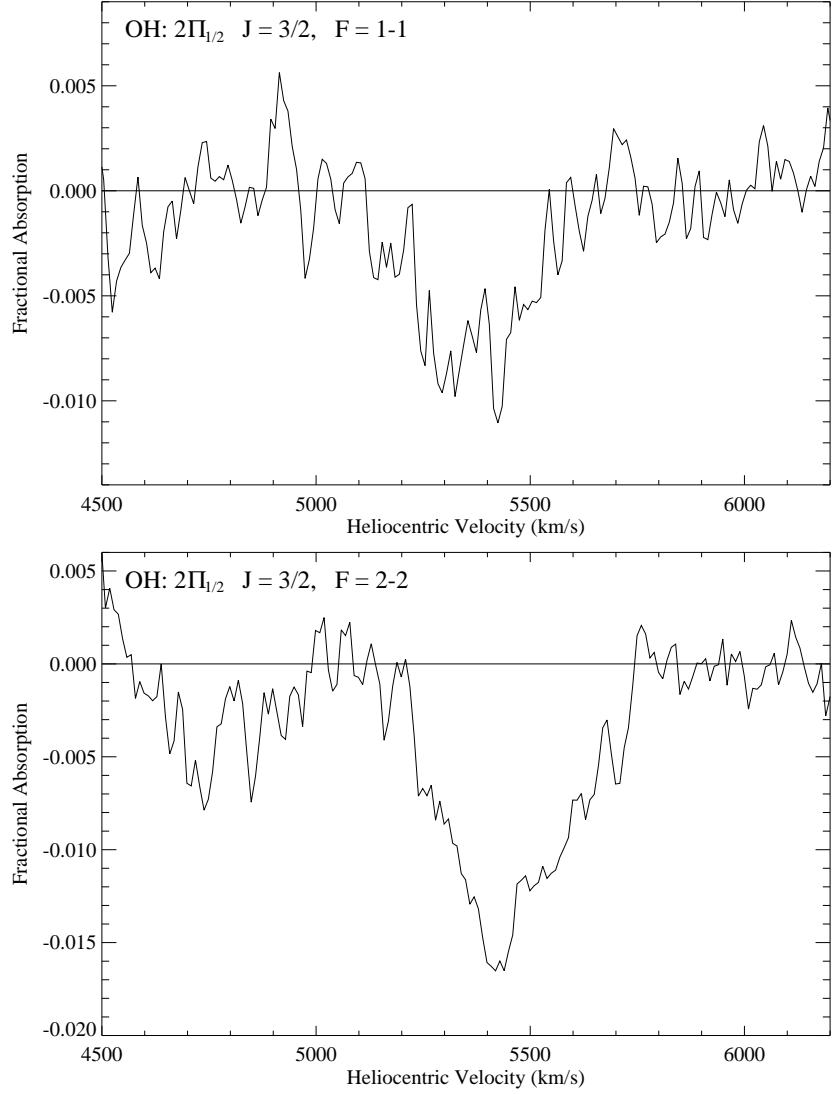}
\caption{First detections of the two $\lambda$4-cm excited OH main-line
transitions, $2\Pi_{1/2}$, J=3/2, F=1-1 and 2-2 (at 7761 and 7820~MHz
respectively). The spectra are plotted with heliocentric velocity as
abscissa.  The velocity resolution is $\sim$30~km\,s$^{-1}$.
\label{f5}}                                                                 
\end{figure}

\begin{figure}
\epsscale{0.8}
\plotone{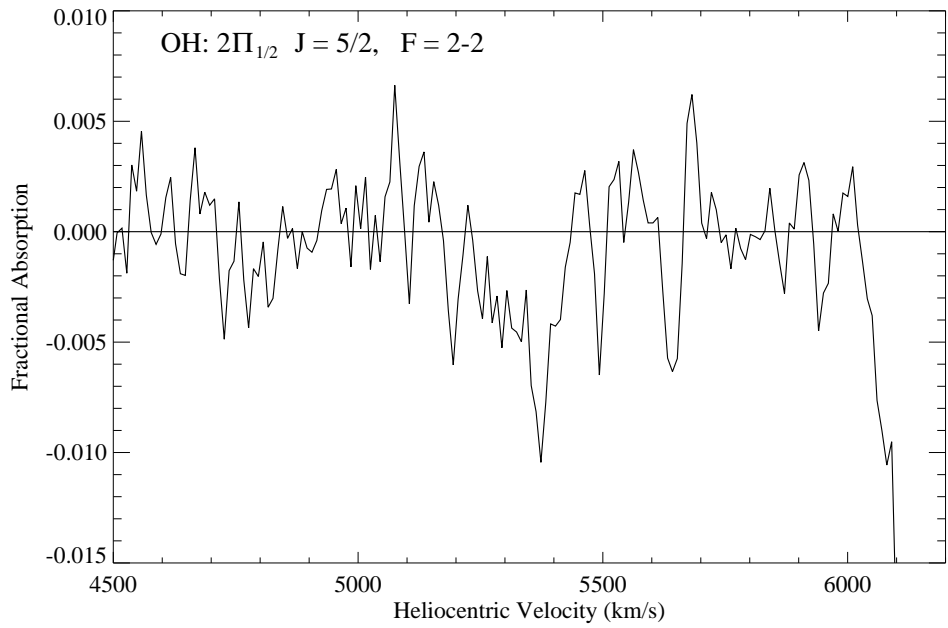}
\caption{A possible detection of the $\lambda$4-cm excited OH transition
$2\Pi_{1/2}$, J=5/2, F=2-2 (at 8135~MHz). The spectrum is plotted
with heliocentric velocity as abscissa.  The velocity resolution is
$\sim$30~km\,s$^{-1}$.
\label{f6}}                                                                 
\end{figure}

The energy levels for the various OH transitions are shown in
Fig.~\ref{OH_energy_diag}.  In Figs.~\ref{f3}--\ref{f6} we present
the $\lambda$6-, 5- and 4-cm $\Lambda$-doublet transitions of the OH
radical seen in absorption against the continuum emission of Arp~220.
The relevant quantum numbers, rest frequencies and other measured line
parameters are presented in Table~2. The $\lambda$6- and 5-cm lines
have been previously detected by \citet{henk87} and \citet{henk86}
respectively.

Contrary to the findings of \citet{henk87}, we find that the 4660-,
4751- \& 4766-MHz lines have intensity ratios closely in agreement with
their expected local thermodynamic equilibrium (LTE) values of 1:2:1.

The $\lambda$5-cm main lines presented here have considerably higher
signal-to-noise ratio than the measurements of \citet{henk86} but show
similar form.  The expected LTE intensity ratios for the 6017-, 6031-,
6035- \& 6049-MHz lines are 1:14:20:1. However, we measure ratios of
1:13:13:$<$0.4.  The two main lines of the $\lambda$5-cm transition
(6031~and 6035~MHz) are blended due to their large velocity widths and
give the appearance of being ``saturated'', as suggested by their
similar optical depths. This may be due to their having high optical
depths, resulting in saturation of the lines against a continuum
component containing of order 10\% of the total flux density of
Arp~220. However, we note that there is a low-velocity component to
(presumably) the 6031-MHz absorption line, seen in Fig.~6 at 6026~MHz,
which if also present for the 6035-MHz line would contribute
significantly at the frequency of the 6031-MHz line. The satellite
lines are seen at much lower signal-to-noise level. However, the
greater strength of the 6017-MHz line relative to that at 6049~MHz is
interesting. A similar effect was also found by \citet{gw83} for four
compact HII regions in our own Galaxy. While they saw the 6017-MHz
lines enhanced above the LTE ratio to the main lines, the 6049-MHz
line was weaker than predicted, and may even sometimes have appeared
in weak emission.

The two $\lambda$4-cm OH lines at 7761~and 7820~MHz (Fig.~\ref{f5}) are
detected for the first time in Arp~220, and show similar velocity widths
to most other molecular species in the galaxy. The expected ratio for
these $2\Pi_{1/2}, J=3/2$ main lines (7761:7820~MHz) is 1:1.8 in thermal
equilibrium, very close to the derived ratio of the peak and integrated
brightnesses for the lines of 1:1.89 (see Table~2). The satellite lines
in this multiplet are expected to be present only at the  1-$\sigma$ level,
and indeed are not detected.  Considering the absorption lines in the
$2\Pi_{1/2}$ ladder of Fig.~\ref{OH_energy_diag} at 4750~and 7820~MHz,
in thermal equilibrium the ratio of their integrated optical depths
would imply an excitation temperature of about 88~K. In a detailed Large
Velocity Gradient (LVG) study of their CO observations, \cite{gre06}
find that in Arp~220, the spectra of low density tracers such as CO (and
OH) can only be fitted with a two-phase molecular ISM with the kinetic
temperature of one being $> 30$~K, while their mm-wave spectra for the
high-density tracer molecules, HCN and CS, indicate kinetic temperatures
of $\sim$50-70~K.

A very tentative detection of the $2\Pi_{1/2}, J=5/2, F=$ 2--2 line of
OH at 8135~MHz is shown in Fig.~\ref{f6}. If confirmed by the addition
of the remaining data to be acquired in this project, this would be at
least twice as strong as expected for an excitation temperature of 88~K.
Further, this line is expected to be only 0.7 times as strong as for the
other $2\Pi_{1/2}, J=5/2$ main line at 8189~MHz. While the 8189~MHz line
is also possibly detected, contrary to expectations this would be at an
even lower signal-to-noise level than for the 8135~MHz transition.

\subsection{{\it Formic Acid (HCOOH)} or {\it ``Heavy'' Hydroxyl
($\,^{18}$OH)}}

\begin{figure}
\epsscale{0.8}
\plotone{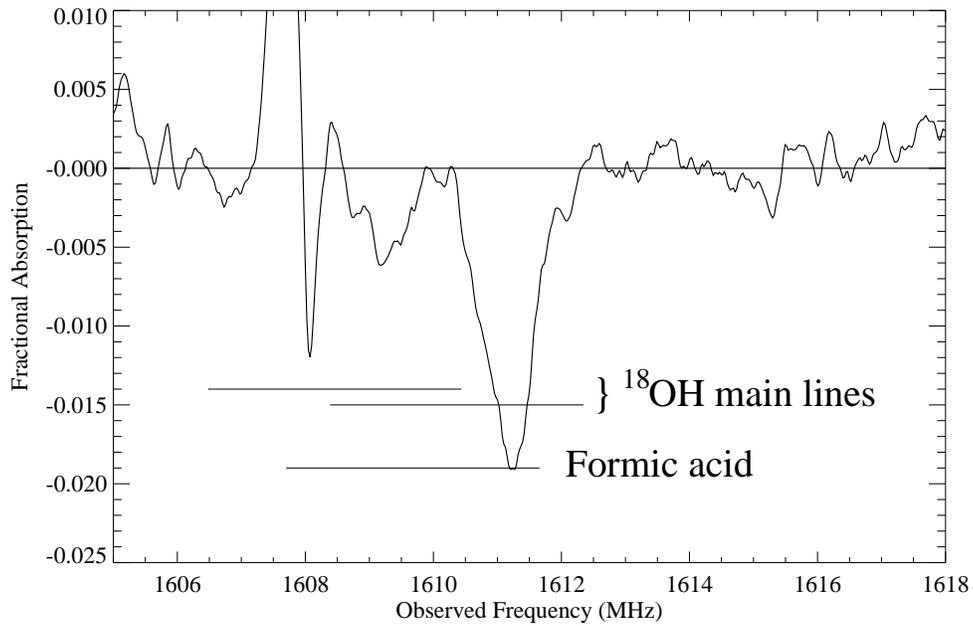}
\caption{This plot shows one (or possibly two) absorption line(s) that could
be either formic acid or $^{18}$OH. The expected location of the formic acid
line and of the two main $^{18}$OH lines are indicated with horizontal bars.
The band is affected by interference from the Glonass system at about
1605~MHz, but the authenticity of the deep absorption feature has been
verified by its non-appearance in the spectra of the bandpass calibrator.
The velocity resolution is $\sim$30~km\,s$^{-1}$.
\label{f7}}                                                                 
\end{figure}

\begin{figure}
\epsscale{0.8}
\plotone{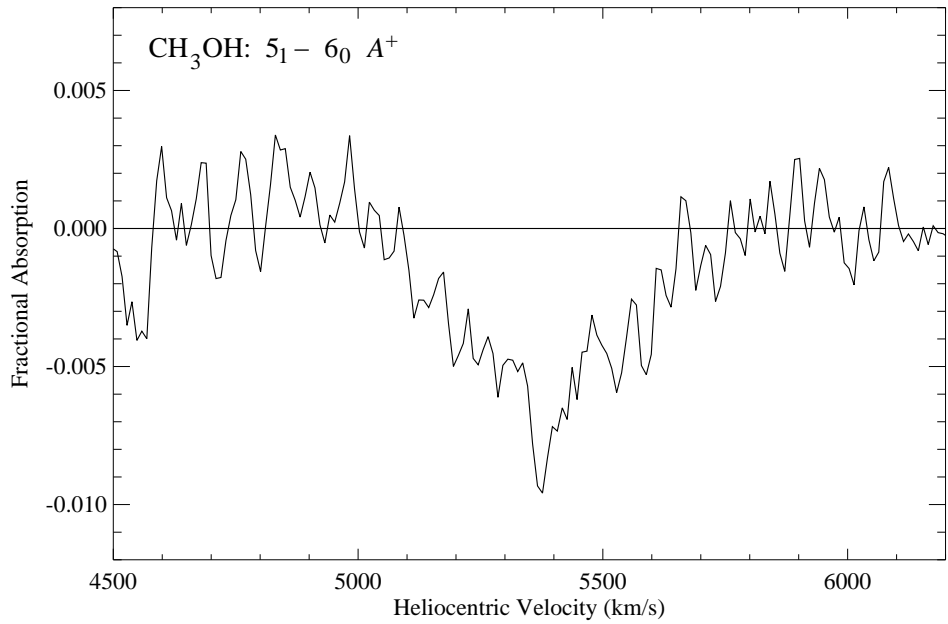}
\caption{A possible detection of the $5_{1}$--$6_{0}$~${\it A}^{+}$
methanol line (at 6668~MHz) in absorption. However, more data is needed
to confirm the detection.  The spectrum is plotted with heliocentric
velocity as abscissa.  The velocity resolution is $\sim$30~km\,s$^{-1}$.
\label{f8}}                                                                 
\end{figure}

In Fig.~\ref{f7}, we present a high signal-to-noise detection of an
absorption line from our L-band spectrum. Despite the presence of
nearby RFI caused by Glonass emissions (at $\sim$1605~MHz), the reality
of this absorption line has been verified by its presence with similar
optical depth in each of 13 individual spectra of Arp~220, but in none
of the spectra of the bandpass calibrator.  As demonstrated by the
horizontal lines in Fig.~\ref{f7}, there is an ambiguity as to the
species responsible for this absorption, which could be either the
1639.5-MHz main line of $^{18}$OH or the 1638.8-MHz line of formic acid
(HCOOH).  Given the prevalence of OH in Arp~220, it would
perhaps not be unreasonable to detect the presence of $^{18}$OH as
well.  However, since formic acid is relevant to the chemical origin of
life, it would be most interesting to resolve this ambiguity.

As is seen in Fig.~\ref{f7}, as well as the strong absorption line
detected near 1611~MHz, an apparently weaker absorption is also seen
near 1609~MHz, close to where the second, 1637.6-MHz, main line of $^{18}$OH
should be found. If this feature were indeed to be real, the ratio of
the peak intensity of the higher frequency absorption to this would be
3.1:1. This is higher than the expected ratio of 1.8:1 were the pair 
to be the main lines of $^{18}$OH \citep{barr64} and in local thermal
equilibrium. The frequencies of the peak depths of the two features are
separated by an amount that would correspond to a rest frequency
separation of about 2.05~MHz were they both to be due to absorption in
Arp~220. The laboratory separation of the $^{18}$OH main lines is
$1.939 \pm 0.003$~MHz \citep{lov86}, and hence given the weakness of
the feature near 1609~MHz, the agreement is considered to be
satisfactory.

If the absorption near 1611~MHz is indeed the higher frequency
component of the $^{18}$OH main lines, then its radial velocity is
closer to that of the peak velocity found for other molecules in
Arp~220 than would be the case were the detection to be of formic acid
(see Table~2 and the horizontal bars displayed in Fig.~\ref{f7}.) However,
even for $^{18}$OH, this velocity of $\sim$5265~km\,s$^{-1}$
is significantly lower than that of the normal molecular peak velocity in
Arp~220. In addition, its width of about 190~km\,s$^{-1}$ is narrower than
for any other species that we detect. Were this absorbing gas to represent
$^{18}$OH, we note that a velocity of 5265~km\,s$^{-1}$ is ``allowed''
in Arp~220 from the HI absorption study of \citet{mun01}.  However, a peak
velocity of 5135~km\,s$^{-1}$, appropriate for the formic acid transition,
has little associated HI absorption. Nevertheless, if the absorption were
to be due to $^{18}$OH at 5265~km\,s$^{-1}$ this would imply a remarkably
high isotopic ratio of $^{18}$OH:$^{16}$OH for a cloud at this velocity,
especially so as this is well away from the peak velocity range of the
OH megamaser line as seen at low angular resolution \citep{baan82}.

Given the proximity of the Glonass RFI, it is difficult to establish
the origins of this absorption feature. However, $^{18}$OH would seem
a more likely identification than formic acid. Confirmation of the
reality of the second, weaker, absorption component near 1609~MHz
would effectively establish this identification. The possibility that
the feature represents absorption in our galaxy from the 1612.231~MHz
satellite OH line has to be very small given the high galactic
latitude of the line-of-sight ($b=53^{\circ}.0$), the implied large
radial velocity ($\sim$+185~km\,s$^{-1}$), and large line width
($\sim$190~km\,s$^{-1}$).

\subsection{\it Methanol (CH$_{3}$OH)}

In Fig.~\ref{f8}, we present a possible detection of the
$5_{1}$--$6_{0}$~${\it A}^{+}$ methanol line in absorption. Although
this line is apparently detected with a signal-to-noise ratio of almost
6:1, and the entire 100-MHz band in which it is observed is basically
RFI-free, the quality of the baseline in this case is rather poor.
We await our remaining observations for confirming the reality of this
line. We note that while the excited-OH absorption line at 7761.7~MHz
(Fig.~\ref{f5}: upper) has a somewhat lower signal-to-noise ratio than
this possible methanol detection, the combination of a better overall
baseline, and the 10.5-$\sigma$ level detection of the associated line at
7820.1 MHz (Fig.~\ref{f5}: lower) yielding the expected LTE intensity
ratio (see Section~\ref{OH_lines}), makes for a more solid detection in
that case.  We can certainly conclude that at the 1~mJy level (5$\sigma$)
no methanol maser emission, such as is commonly detected for this
transition from regions of massive-star formation in our own Galaxy,
is seen in Arp~220.

Assuming the reality of the methanol absorption line, 
an excitation temperature for the transition of T$_{ex} = 20$\,K, and
an Einstein coefficient of $A = 0.1532 \times 10^{-8}$~s$^{-1}$
\citep{cr93}, and a covering factor of unity, in LTE the column density
of the lower energy level, $N_{6,0}$ is;

\begin{equation}
N_{6,0} = 1.34 \times 10^{15} \int \tau \delta v~~{\rm cm}^{-2}
\end{equation}

where the integral $\int \tau \delta v$ is expressed in units of
km\,s$^{-1}$.

An approximation for the total column density of methanol can be
obtained assuming a Boltzmann distribution for the methanol
energy-level population. If the relative abundances of the
A and E species of this molecule is taken to be 1:1 based on their
having a relatively small ground state energy difference, then the
total column density of methanol molecules is;

\begin{equation}
N = N_{6,0}~2/13~Q(T_{ex})~e^{\frac{E_{6,0}}{kT_{ex}}}
\end{equation}

where Q(T$_{ex}$) is the partition function, taken to be 39.8 (Cragg,
D., quoted in Houghton~\& Whiteoak 1995), and E$_{6,0}$ is the energy
above the ground state of the 6$_{0}$A$^{+}$ level. Thus, the presence
of a 6668-MHz methanol absorption line in Arp~220 would imply a total
column density for this molecule of $N_{\rm CH_{3}OH} \sim 2.5 \times
10^{17}$~cm$^{-2}$.

\section{Concluding Remarks}

The Arecibo Arp~220 Spectral Census is an on-going project and a
detailed analysis will be presented in a separate paper, as will the
final spectra from the completed project.  We have presented here the
discovery of methanimine in this galaxy and the detection of three
previously unseen cm-wavelength transitions of HCN.  We have also
observed, for the first time in Arp~220, an absorption line that may
be either formic acid or $^{18}$OH, two (possibly three) $\lambda$4-cm
transitions of excited OH, and what may be the first extragalactic
detection of methanol.

The discovery of high abundances of ``pre-biotic'' molecules, such
as methanimine, HCN and possibly formic acid in Arp~220 raises the
possibility that other ULIRGs might contain similarly high abundances
of such molecules, which could be detectable with the Arecibo 305-m
telescope.

\acknowledgments

We thank Paul Goldsmith (JPL), Sven Thorwirth and Karl Menten (MPIfR)
for fruitful discussions, and an anonymous referee for a number of
useful suggestions which considerably improved the paper. The Arecibo
Observatory is a part of the National Astronomy and Ionosphere Center
(NAIC) operated by Cornell University under a cooperative agreement
with the National Science Foundation (NSF).

{\it Facilities:} \facility{Arecibo (L-wide, S-high, C, C-high, X)}

\clearpage

%% Tables should be submitted one per page, so put a \clearpage before
%% each one.

%% Two options are available to the author for producing tables:  the
%% deluxetable environment provided by the AASTeX package or the LaTeX

\clearpage

\begin{deluxetable}{ccccccc}
\tabletypesize{\scriptsize}
%%\rotate
\tablewidth{0pt}
\tablecaption{Derived parameters for the emission multiplet of methanimine
(CH$_{2}$NH).}

\tablehead{
   \colhead{Molecule} &
   \colhead{Transition} &
   \colhead{\vtop{\hbox{Rest frequency\strut}\hbox{\nn\nn\nn\np(MHz)\strut}}} &
   \colhead{\vtop{\hbox{\nn\np$S_p$\strut}\hbox{(mJy)\strut}}} &
   \colhead{\vtop{\hbox{\nn\nn$\sigma$\strut}\hbox{(mJy)\strut}}} &
   \colhead{\vtop{\hbox{\nn\nn\nn$v_p$\strut}\hbox{(km\,s$^{-1}$)\strut}}} &
   \colhead{\vtop{\hbox{\np{FWHM}\strut}\hbox{(km\,s$^{-1}$)\strut}}} 
}
\startdata
Methanimine (CH$_{2}$NH) & $1_{10}$--$1_{11}, \Delta F=0, \pm 1 $ &
   5289.813$^{1}$ & 3.5 & 0.22 & 5362 & 270 \\
\enddata
\tablenotetext{\relax}{\noindent The subscript $p$ stands for ``peak''. \\ 
$^1$ The strongest of the six lines of this multiplet has this frequency in
laboratory measurements.}
\end{deluxetable}

\begin{deluxetable}{lrrrrccc}
\tabletypesize{\scriptsize}
%%\rotate
\tablewidth{0pt}
\tablecaption{Derived parameters from the molecular absorption lines.}

\tablehead{
   \colhead{Molecule} &
   \colhead{Transition} &
   \colhead{\vtop{\hbox{Rest frequency\strut}\hbox{\nn\nn\nn\np(MHz)\strut}}} &
   \colhead{\vtop{\hbox{\nn\np$\tau_p$\strut}}} &
   \colhead{\vtop{\hbox{\nn\np$\sigma_{\tau}$\strut}}} &
   \colhead{\vtop{\hbox{\nn\nn\nn$v_p$\strut}\hbox{(km\,s$^{-1}$)\strut}}} &
   \colhead{\vtop{\hbox{\np{FWHM}\strut}\hbox{(km\,s$^{-1}$)\strut}}} &
   \colhead{\vtop{\hbox{\nn$\int \tau dV$\strut}\hbox{(km\,s$^{-1}$)\strut}}}
}
\startdata
Hydrogen cyanide (HCN) & $v_{2}=1, \Delta J=0, J=2$ &
   1346.7650\nn\nn & $<0.0025$ & 0.00085 & \nodata & \nodata & \nodata \\
 & $J=4$ &
   4488.4718\nn\nn & 0.016\nn & 0.0009\nn & 5407 & 363 & \nn5.22 \\
 & $J=5$ &
   6731.9098\nn\nn & 0.0205 & 0.001\nn\nn & 5404 & 330 & \nn5.02 \\
 & $J=6$ &
   9423.3338\nn\nn & 0.0268 & 0.002\nn\nn & 5398 & 202 & \nn6.14 \\ 
Hydroxyl radical (OH) & $2\Pi_{1/2}, J=1/2, F=$ 0--1 &
   4660.242\nn\nn\nn & 0.0254 & 0.00082 & 5425 & 300 & \nn7.99 \\ 
 & $F=$ 1--1 &
   4750.656\nn\nn\nn & 0.0549 & 0.00080 & 5429 & 287 & 15.91 \\
 & $F=$ 1--0 &
   4765.562\nn\nn\nn & 0.0295 & 0.00080 & 5447 & 281 & \nn8.41 \\
 & $2\Pi_{3/2}, J=5/2, F=$ 2--3 &
   6016.746$^{2}$\nn\nn & 0.0086 & 0.0016\nn & 5313 & \nodata & \nodata \\
 & $F=$ 2--2 & 6030.747$^{1}$\nn\nn & 0.11\nn\nn & 0.0016\nn & 5425 & 480 &
   56.20 \\
 & $F=$ 3--3 & 6035.092$^{1}$\nn\nn &  &  &  &  &  \\ 
 & $F=$ 3--2 & 6049.084\nn\nn\nn & $<0.0033$ & \nodata\nn & \nodata & \nodata &
   \nodata \\
 & $2\Pi_{1/2}, J=3/2, F=$ 1--1 &
   7761.747\nn\nn\nn & 0.0087 & 0.0018\nn & 5359 & 296 & \nn2.62 \\
 & $F=$ 2--2 &
   7820.125\nn\nn\nn & 0.0162 & 0.00153 & 5432 & 321 & \nn5.04 \\
 & $2\Pi_{1/2}, J=5/2, F=$ 2--2 &
   8135.870$^{2}$\nn\nn & $<0.01$\nn\nn & 0.0021\nn & \nodata & \nodata &
   \nodata \\
$^{18}$OH {\it or\/} & $2\Pi_{3/2}, J=3/2, F=$ 2--2 &
   1639.503\nn\nn\nn & 0.019\nn & 0.00137 & 5265 & \llap{$\sim$}192 &
   ---\rlap{$^{3}$} \\
\nn\nn Formic acid (HCOOH) & 1(1,0)--1(1,1) &
   {\it or\/} 1638.805\nn\nn\nn &  & & 5135 &  & \\
Methanol (CH$_{3}$OH) & $5_{1}$--$6_{0} {\it A}^{+}$ &
   6668.5192\nn\nn & 0.0077 & 0.0013\nn & 5384 & 360 & \nn2.53 \\
\enddata
\tablenotetext{\relax}{\noindent The subscript $p$ stands for ``peak''. \\ 
$^1$ Since the two main lines $F=$ 2-2 and 3-3 are blended, values
presented here are for the two lines combined together. \\
$^2$ For these low signal-to-noise detections some of the parameters were
not estimated. \\
$^3$ Since the identification of this absorption feature is ambiguous,
an optical-depth integral was not estimated for it.}
\end{deluxetable}

\end{document}